\documentclass[%
 reprint,
superscriptaddress,
 amsmath,amssymb,
 aps,prl,
]{revtex4-2}

\usepackage{graphicx}% Include figure files
\usepackage{dcolumn}% Align table columns on decimal point
\usepackage{bm}% bold math
\usepackage{subcaption}
\usepackage{color}
\RequirePackage[colorlinks=true
,urlcolor=blue
,anchorcolor=blue
,citecolor=blue
,filecolor=blue
,linkcolor=blue
,menucolor=blue
,linktocpage=true
,pdfproducer=medialab
,pdfa=true
]{hyperref}
\begin{document}
\title{Dark matter candidate from torsion}
\author{\'Alvaro de la Cruz Dombriz}
%\email{alvaro.dombriz@usal.es}
\affiliation{Departamento de F\'isica Fundamental, Universidad de Salamanca, E-37008 Salamanca, Spain}
\affiliation{Cosmology and Gravity Group, Department of Mathematics and Applied Mathematics, University of Cape Town, Rondebosch 7701, Cape Town, South Africa}
\author{Francisco Jos\'{e} Maldonado Torralba}
%\email{fmaldo01@ucm.es}
\affiliation{Cosmology and Gravity Group, Department of Mathematics and Applied Mathematics, University of Cape Town, Rondebosch 7701, Cape Town, South Africa}
\affiliation{Institute of Theoretical Astrophysics, University of Oslo, P.O. Box 1029 Blindern, N-0315 Oslo, Norway}
\author{David F. Mota}
%\email{d.f.mota@astro.uio.no}
\affiliation{Institute of Theoretical Astrophysics, University of Oslo, P.O. Box 1029 Blindern, N-0315 Oslo, Norway}
\date{\today}% It is always \today, today,
\begin{abstract}
The stable pseudo-scalar degree of freedom of the quadratic Poincar\'e Gauge theory of gravity is shown to be a suitable dark matter candidate. We find the parameter space of the theory which can account for all the predicted cold dark matter, and constrain such parameters with astrophysical observations.
\end{abstract}
 %                             %display desired
\maketitle
%
%\tableofcontents
%
\textit{Introduction.}--- 
%When facing a novel measure in Nature which cannot be described with the current theoretical framework, we are entitled to follow two different approaches: either there is some elusive form of matter or energy that is affecting such a measure, or the theory is incomplete because it is no longer validated by experiment. 
%At the moment we know at least two 
Several cosmological and astrophysical phenomena cannot be explained resorting to General Relativity (GR) and the matter content of the Standard Model of Particles (SM). For instance, the present day accelerated expansion of the Universe \cite{Riess:1998cb}, and the rotational curves of galaxies do not fit the GR predictions with baryonic matter \cite{Sahni:2004ai}. 
These problems are solved assuming GR as the correct gravitational framework and modelling the accelerated expansion with a new form of energy (\emph{dark energy}) encoded in a cosmological constant $\Lambda$ \cite{Peebles:2002gy}, and the rotation curves by adding a new form of matter
%, which interacts weakly and gravitationally. This kind of matter is 
known as \emph{cold dark matter} (CDM). 
%Therefore, the preferred cosmological model is known as $\Lambda$CDM \cite{weinberg2008cosmology}. 
The mentioned approach suffers from some shortcomings: the theoretical value of $\Lambda$ exceeds the observational value by 120 orders of magnitude \cite{Weinberg:1988cp}, and there is no direct nor conclusive evidence of the CDM particles 
%se weakly interacting particles, we just know there are some proposals for which their effect is compatible with the current measures
besides their gravitational effects at astrophysical scales \cite{Bertone:2004pz}. 
%Furthermore, even if we consider $\Lambda$CDM as the description of the cosmology, there exists a tension between local and late-time measurements of the Hubble parameter, which accounts for the rate of expansion of the Universe \cite{Riess:2019qba}.

%Consequently, it is physically relevant to study 
Another approach is to investigate if both the accelerated expansion and the rotation curves can be understood from modifications of GR. Any Lorentz invariant four-dimensional local modification of the Einstein-Hilbert action of GR necessarily introduces new degrees of freedom (d.o.f.s). Then, we can ask if such d.o.f.s can be used to model dark matter and/or dark energy. This line of thought has been explored in the literature for some modifications of GR (see \cite{Cembranos:2008gj,Clifton:2011jh,Amendola:2016saw}).
%, such as $f(R)$ \cite{delaCruz-Dombriz:2006kob,Amendola:2006we,Cognola:2007zu,Hu:2007nk,Cembranos:2008gj}, k-essence \cite{Armendariz-Picon:2000nqq}, vector-tensor theories \cite{BeltranJimenez:2008iye}, or scalar-tensor-vector gravity \cite{Moffat:2005si}, amongst many others attempts (see \cite{Copeland:2006wr,Clifton:2011jh,Joyce:2014kja,Bull:2015stt,Amendola:2016saw}).

%When exploring modifications of GR we have to be careful with the fact that the new propagating degrees of freedom may have a pathological behaviour, hence we need to make sure that the modified theory is stable. Also, it needs to be compatible with the current experimental data. As a matter of fact, the detection of the gravitational wave GW170817 and its electromagnetic counterpart from a binary neutron star system \cite{LIGOScientific:2017vwq,LIGOScientific:2017ync}, discarded many modified gravity theories \cite{Ezquiaga:2017ekz,Baker:2017hug,Creminelli:2017sry,Sakstein:2017xjx}. Lastly, but not less important, we need to give a theoretical justification of the introduction of new degrees of freedom to the gravitational theory.

%Given the above reasons, 
We shall work with the Poincar\'e Gauge modification of GR. This theory arises naturally when promoting the global Poincar\'e symmetry to a local one, following the gauge procedure. Then, the torsion tensor $T^\rho\,_{\mu\nu}$, which is the antisymmetric part of the spacetime connection $\Gamma^\rho\,_{\mu\nu}$, can be identified as the gauge field strength of spacetime translations $T(4)$. Also, the Riemann tensor $R^\mu\,_{\nu\rho\sigma}$ is given as the gauge field strength of the homogeneous Lorentz group $SO(1,3)$. As usual in Yang-Mills theories, one can construct the Lagrangian considering up to quadratic terms in the field strengths,
\begin{eqnarray}
\mathcal{L}_{{\rm PGG}}\,=\,&&a_{0}R+a_{1}T_{\mu\nu\rho}T^{\mu\nu\rho}+a_{2}T_{\mu\nu\rho}T^{\nu\rho\mu}+a_{3}T_{\mu}T^{\mu}
\nonumber
\\
&&+b_{1}R^{2}+b_{2}R_{\mu\nu\rho\sigma}R^{\mu\nu\rho\sigma}+b_{3}R_{\mu\nu\rho\sigma}R^{\rho\sigma\mu\nu}
\nonumber
\\
&&+b_{4}R_{\mu\nu\rho\sigma}R^{\mu\rho\nu\sigma}+b_{5}R_{\mu\nu}R^{\mu\nu}+b_{6}R_{\mu\nu}R^{\nu\mu},
\label{quad:PG}
\end{eqnarray}
which is known as Poincar\'e Gauge Gravity (PGG) \cite{Blagojevic:2013xpa,ponomarev2017gauge}. %First introduced by Sciama \cite{sciama1962analogy} and Kibble \cite{Kibble:1961ba}, and then formalised by Hayashi \cite{Hayashi:1968hc} and Hehl \emph{et al.} \cite{Hehl:1976kj}, this theory has been widely studied in the literature. %Paradigmatic examples include the study of singularities~\cite{Stewart:1973ux,Trautman:1973wy,Cembranos:2016xqx}, the Birkhoff's theorem \cite{Neville:1979fk,Rauch:1981tva,delaCruz-Dombriz:2018vzn}, finding novel exact solutions \cite{Bakler:1984cq,Obukhov:1987tz,Blagojevic:2015zma,Cembranos:2016gdt,Obukhov:2019fti}, cosmology \cite{Kerlick:1975tr,Yo:2006qs,Shie:2008ms,Chen:2009at,Baekler:2010fr,Ho:2015ulu}, the motion of particles~\cite{Hehl:2013qga,Cembranos:2018ipn} and the analysis of the stability of the theory~\cite{Sezgin:1979zf,Sezgin:1981xs,Chern:1992um,Yo:1999ex,Yo:2001sy,BorislavovVasilev:2017xpl,Jimenez:2019qjc}. 
We know from \cite{Hayashi:1980qp} that, apart from the usual graviton, the matter content of PGG consists of two massive spin-2, two massive spin-1 and two spin-0 fields. In \cite{Yo:1999ex,Yo:2001sy,Jimenez:2019qjc} authors found that the only modes that could propagate safely were the two spin-0 with different parity.

In this letter we study the pseudo-scalar mode and show that it behaves like an axion-like particle (ALP). Then, we find the experimental constraints from the interaction of such a mode with the SM sector. Finally, we give the conditions for which the pseudo-scalar mode of PGG can describe the predicted CDM density.\\

\textit{The pseudo-scalar mode.}--- By considering $b_3=b_1+b_2$, $b_4=-4b_2$, $b_5=0$, and $b_6=-4b_1$ in \eqref{quad:PG}, we get the PGG Lagrangian whose only propagating mode is the pseudo-scalar, 
\begin{equation}
\label{eq:Holst}
    \mathcal{L}=a_0\mathring{R}+\frac{1}{2}m_T^2 T^2+\frac{1}{2}m_S^2S^2
    +\alpha \mathcal{H}^2 .
\end{equation}
Here $\mathcal{H}\equiv\epsilon^{\mu\nu\rho\sigma} R_{\mu\nu\rho\sigma}$ is the Holst term \cite{Hojman:1980kv, Holst:1995pc},
%\footnote{Although this term is commonly known as the Holst term, due to S. Holst's article in 1995 \cite{Holst:1995pc}, it was first introduced by R. Hojman {\emph{et. al.}} in 1980 in the context of torsion gravity \cite{Hojman:1980kv}.}
the $\mathring{\,}$ denotes quantities calculated with respect to the Levi-Civita (LC) connection, $\alpha=-b_2/4$, $m_{T}^{2}=-\frac{2}{3}\big(2a_{0}-2a_{1}+a_{2}-3a_{3}\big)$, and $m_{S}^{2}=\frac{1}{12}\big(a_{0}-4a_{1}-4a_{2}\big)$. Also, $T_{\mu}\equiv T^{\nu}\,_{\mu\nu}$ and $S_{\mu} \equiv \epsilon_{\mu\nu\rho\sigma}T^{\nu\rho\sigma}$ represent the trace vector and axial vector of the torsion tensor respectively. 
%Finally, given the choice of parameters, $b_1$ is the coupling constant of the Gauss-Bonnet term, which we have omitted since it is a total derivative in 4 dimensions. 

Introducing an auxiliary field $\hat{\phi}$, one rewrites \eqref{eq:Holst} as
\begin{equation}
\label{eq:Holst2}
    \mathcal{L}=a_0\mathring{R}+\frac12m_T^2 T^2+\frac12m_S^2S^2-\alpha\hat{\phi}^2+2\alpha\hat{\phi}\epsilon^{\mu\nu\rho\sigma} R_{\mu\nu\rho\sigma}.
\end{equation}
%The resulting Lagrangian corresponds to the addition of a Holst term where the Barbero-Immirzi parameter acts as a pseudo-scalar field.
The massless theory with $m_ T^2=m_S^2=0$ and without the $\hat{\phi}^2$ potential was first considered in \cite{castellani1991supergravity}. Independently, such a theory was proposed as an extension of GR inspired by Loop Quantum Gravity \cite{Taveras:2008yf,Calcagni:2009xz,TorresGomez:2008fj,Mercuri:2009zi}.

From \eqref{eq:Holst2}, the effective action of the pseudo-scalar can be constructed. First, we need to take into account the couplings to matter for such a Lagrangian, which in the minimal coupling prescription is just given by the coupling of the axial torsion vector $S_{\mu}$ to the axial current of fermions $\Delta_\mu$ \cite{Shapiro:2001rz}. Secondly, by resorting to the field equations for the trace and axial vectors we can isolate such vectors with respect to the rest of variables. Plugging that back to \eqref{eq:Holst2} we find (see \cite{Jimenez:2019qjc} for details):
\begin{equation}
\label{Holst:Poincare}
    \mathcal{L}=\frac{1}{2}a_0\mathring{R}-\frac{\left(2\alpha\partial_{\mu}\hat{\phi}+\Delta_{\mu}\right)^{2}}{2\left[m_{S}^{2}-\left(\frac{4\alpha\hat{\phi}}{3m_{T}}\right)^{2}\right]}-\alpha\hat{\phi}^{2}.
\end{equation}
Due to the characteristic derivative coupling to a current, this pseudo-scalar d.o.f. $\hat{\phi}$ can be identified as an ALP \cite{ParticleDataGroup:2020ssz}. 
%This theory has conceptual and phenomenological differences with respect to the mentioned massless cases in the literature, for which the effective theory was calculated in \cite{TorresGomez:2008fj}. On the one hand, in those proposals there is no potential term of the pseudo-scalar, although it can be added \emph{ad hoc}. This is due to the fact that the considered action therein is a minimal extension of Einstein-Cartan Gravity, which consists in the Ricci scalar of the total connection plus the pseudo-scalar field coupled to the Holst term. In \eqref{Holst:Poincare} the potential term is present because the original Lagrangian contains the Holst square. As seen in \eqref{eq:Holst}, this theory can be obtained by a suitable choice of parameters in the quadratic Poincar\'e Gauge Lagrangian. Therefore, we are not extending any theory, we are studying a stable propagating degree of freedom of one. Finally, in the aforementioned models the mass is fixed by the way the potential term is introduced, and in many cases is fixed. On the contrary, as we will see, in our case the mass of the predicted particle depends on the parameters, and consequently it is not fixed by the theory.

At tree-level the pseudo-scalar couples to fermionic particles only  through a derivative coupling. When considering quantum corrections, new interactions arise due to the so-called axial anomaly \cite{Marsh:2015xka}. In the case of a curved spacetime, such an anomaly can be expressed as \cite{Bertlmann:1996xk}
\begin{equation}
\label{anomaly:1}
    \left.\partial_{\mu}\Delta^{\mu}\right|_{{\rm anomaly}}=\frac{e^2}{32\pi^2}F^{\mu\nu}\tilde{F}_{\mu\nu}-\frac{1}{384\pi^{2}}\mathcal{K}_2,
\end{equation}
where $e$ is the electron charge, $F_{\mu\nu}$ denotes the usual electromagnetic tensor, $\tilde{F}_{\mu\nu}=\frac{1}{2}\epsilon_{\mu\nu\rho\sigma}F^{\rho\sigma}$ refers to its dual, and $\mathcal{K}_2=\epsilon^{\mu\nu\rho\sigma}\mathring{R}^{\alpha\beta}\,_{\mu\nu}\mathring{R}_{\alpha\beta\rho\sigma}$ is known as the Chern–Pontryagin scalar.

In order to find the couplings of the pseudo-scalar, we first canonically normalise the Lagrangian by making the transformation
\begin{equation}
\label{canonical}
    \phi=\frac{2\alpha}{\sqrt{m_{S}^{2}}}\int\frac{{\rm d}\hat{\phi}}{\sqrt{1-\left(\frac{4\alpha\hat{\phi}}{3m_{T}m_{S}}\right)^{2}}}.
\end{equation}
Depending on the signs of $m_T^2$ and $m_S^2$, the normalised effective Lagrangian would be different. In all the relevant cases, after having taken into account the axial anomaly by integrating by parts the anomalous part of the derivative coupling term in the Lagrangian, the Lagrangian for the pseudo-scalar becomes
\begin{eqnarray}
\label{phi:1loop}
    \mathcal{L}_{\phi}&=&\frac{1}{2}a_0\mathring{R}-\frac{1}{2}\partial_{\mu}\phi\partial^{\mu}\phi-\frac{1}{\alpha}\left(\frac{3m_{T}m_{S}}{4}\right)^{2}H_{1}\left(\phi\right)
    \nonumber
    \\
    &&-\frac{1}{m_{S}}H_{2}\left(\phi\right)\Delta_{{\rm N}}^{\mu}\partial_{\mu}\phi+\frac{e^{2}}{32\pi^{2}m_{S}}H_{3}\left(\phi\right)F^{\mu\nu}\tilde{F}_{\mu\nu}
    \nonumber
    \\
    &&-\frac{1}{384\pi^{2}m_{S}}H_{3}\left(\phi\right)\mathcal{K}_2-\frac{1}{2m_{S}^{2}}H_{2}^{2}\left(\phi\right)\Delta^{\mu}\Delta_{\mu},
\end{eqnarray}
where $\Delta^\mu_{\rm N}$ is the non-anomalous part of the axial current.
The meaning of functions $H_{1,2,3}$ depending on the sign of the $\left\{m_S,m_T\right\}$ parameters has been summarised in Table \ref{tab:table1}.
\begin{table}
\caption{\label{tab:table1}%
Functions involved in Eq. \eqref{phi:1loop} depend on the sign of $m^2_T$ and $m^2_S$. The case $m^2_S<0,m^2_T>0$ has not been considered since it renders the pseudo-scalar to be unstable \cite{Jimenez:2019qjc}. Also, the case $m^2_S,m^2_T<0$ is not given below since it lacks a well-defined weak-field limit.
}
\begin{tabular}{|c||c|c|}
\cline{2-3} 
\multicolumn{1}{c||}{} & $m^2_S,m^2_T>0$ & $m^2_S>0,m^2_T<0$\tabularnewline
\hline 
\hline
$H_{1}$ & $\sin^{2}\left(\frac{2\phi}{3m_{T}}\right)$ & $-\sinh^{2}\left(\frac{2\phi}{3\left|m_{T}\right|}\right)$\tabularnewline
\hline 
$H_{2}$ & $\cos^{-1}\left(\frac{2\phi}{3m_{T}}\right)$ & $\cosh^{-1}\left(\frac{2\phi}{3\left|m_{T}\right|}\right)$\tabularnewline
\hline 
$H_{3}$ & $\int\frac{{\rm d}\phi}{\cos\left(\frac{2\phi}{3m_{T}}\right)}$ & $\int\frac{{\rm d}\phi}{\cosh\left(\frac{2\phi}{3\left|m_{T}\right|}\right)}$\tabularnewline
\hline 
\end{tabular}
\end{table}
For all the cases presented in Table \ref{tab:table1}, the weak field limit $\phi\ll m_T$ for \eqref{phi:1loop} renders 
\begin{eqnarray}
\label{phi:weak}
\mathcal{L}_{{\rm weak}\;\phi}&=&\frac{1}{2}a_0\mathring{R}-\frac{1}{2}\partial_{\mu}\phi\partial^{\mu}\phi-\frac{m_{S}^{2}}{4\alpha}\phi^{2}-\frac{1}{m_{S}}\Delta_{\rm N}^{\mu}\partial_{\mu}\phi
\nonumber
\\
&&+\frac{e^{2}}{32\pi^{2}m_{S}}\phi F^{\mu\nu}\tilde{F}_{\mu\nu}-\frac{\phi}{384\pi^{2}m_{S}}\mathcal{K}_2
\nonumber
\\
&&-\frac{1}{2m_{S}^{2}}\left[1+\left(\frac{2\phi}{3m_{T}}\right)^{2}\right]\Delta^{\mu}\Delta_{\mu}\,,
\end{eqnarray}
which has the form of the usual ALP Lagrangian plus a four-fermion contact interaction $\Delta^{\mu}\Delta_{\mu}$ and the Chern-Simons term $\mathcal{K}_2$. Let us note that for a bounded field $\phi$ we can always choose a value of $m_T$ such that the weak-field approximation is valid. Finally, it can be observed that the predicted mass of the ALP given by the quadratic potential term in \eqref{phi:weak} is $m_{\phi}=\frac{m_{S}^{2}}{2\alpha}$. \\

\textit{Experimental constraints.}--- Given those four interactions of the pseudo-scalar with the SM sector, we can set constraints from experiments. Four different kinds of constraints are considered herein: 1) the ones from the axial-axial interaction, \emph{i.e.}, four-fermion contact interaction, 2) the ones from the Chern–Pontryagin coupling to the pseudo-scalar field, 3) the ones deriving from the coupling with the electromagnetic sector, and 4) the ones from the coupling to the axial current. 
%All these constraints will be presented in the $m_S-\alpha$ parameter space. As we have seen, 
Since the weak-field approximation can be met by tuning $m_T$, we shall use this parameter to set the constraints.

1) The four-fermion contact interaction is constrained by particle-physics observables that would be affected by the addition of such an interaction. Paradigmatic examples of such experiments include HERA, LEP, and the Tevatron. We shall use the constraint set by a global analysis of the results of the aforementioned experiments in reference \cite{Zarnecki:1999je}. We have focused on those ones coming from assuming the exchange of purely axial-vector couplings, which is the case of the contact interaction induced by torsion. Authors in \cite{Zarnecki:1999je} found that the coefficient regulating the contact interaction should be lower than $0.055\;\rm{TeV}^{-2}$. Since such a limit must be true for $\phi=0$, we can find a lower bound $m_{S}>0.166\mathrm{~TeV}$.

2) The Chern-Simons modification of GR is an extension based on the addition of the Chern–Pontryagin term coupled to a scalar field \cite{Alexander:2009tp}. Such a theory is inspired by either the aforementioned gravitational anomaly, or String Theory, or Loop Quantum Gravity. The fact that the gravitational coupling term induced by the anomaly is part of a well-known modified gravity theory, allows us to use the constraints already set in the literature. Such constraints are based on gravitational-wave measures \cite{Jung:2020aem}, binary pulsars \cite{Yunes:2009hc}, and frame-dragging effects \cite{Ali-Haimoud:2011zme}. Nevertheless they are really mild constraints when compared to the ones already obtained from the four-fermion contact interaction. In fact, the constraints on the parameter mediating the interaction, in our case $\left(384\pi m_{S}\right)^{-1}$, are of the order of $10^4\:\rm{km}$, which in natural units translates to $0.507\:\rm{eV}^{-1}$. This gives a lower limit on $m_S$ of $1.63\:\rm{meV}$. Hence, it is clear that even with near future surveys on the mentioned measures, the constraints obtained from the contact interaction would be much stronger. 

3) The coupling of ALPs with the electromagnetic sector, \emph{i.e.} $g_{\phi\gamma\gamma}\phi F^{\mu\nu}\tilde{F}_{\mu\nu}$, has motivated most of the experimental searches for these kinds of particles, since such an interaction predicts the interconversion with photons in the presence of a background magnetic field \cite{Chadha-Day:2021szb}. The effects of this conversion can in principle be measured by several telescopes and ground-based experiments. Paradigmatic examples include (see \cite{Semertzidis:2021rxs}):

\underline{Helioscopes}: These kinds of constraints are based on the assumption that the ALP is a constituent of our galactic halo. Hence, the Sun would be able to produce them, and the photons produced by passing a magnetic field would be detectable on Earth on the X-ray region. Following this reasoning, the most stringent values have been provided by the CERN experiment CAST, which gives an upper bound of the ALP-photon  coupling $g_{\phi\gamma\gamma}$ of $6.6\cdot 10^{-11}\:{\rm{GeV}}^{-1}$ for a mass $m_{\phi}\lesssim0.02\:{\rm eV}$.
   
\underline{Haloscopes}: These detectors are designed to measure microwave-photon signals from axions in our galactic halo. They are able to set the best constraints on the microwave range. Some experiments under this classification include RBF-UF, ADMX, HAYSTAC, CAPP, and for low-mass searches, ABRACADABRA and SHAFT.
%    
%    \item Light-shinning-through-wall experiments (LSW): these experiments aim to measure the photon to ALP to photon ($\gamma\rightarrow{\rm ALP}\rightarrow\gamma$) process. The fact that it has not been observed places constraints on the electromagnetic coupling of the ALP. Experiments of this kind include OSQAR \cite{OSQAR:2015qdv} and ALPS \cite{Ehret:2010mh}.  
%

\underline{Astrophysical measures}: here we include the astrophysical observations that would be affected by ALPs. The study of globular clusters gives one of the strongest constraints for the large-mass range. The number count of Horizontal Branch stars (HB), which can produce axions by the Primakoff process, unlike the red giants, which are not affected by these losses, gives a bound of $6.6\cdot 10^{-11}\:{\rm{GeV}}^{-1}$ to the ALP-photon coupling in a wide mass range \cite{Ayala:2014pea}. For small masses, the measure of X-ray sources with Chandra, gives one of the most stringent constraints. Also, the study of gamma rays from SN 1987A, and the active galactic nuclei of AGN PKS 2155-304 and NGC 1275 (by the HESS and Fermi-LAT collaborations respectively), gives comparable constraints. Finally, for large masses the best bounds are given by the spectroscopic observations of the dwarf spheroidal galaxy Leo T using the MUSE survey, in order to find ALP radiative decay, and of galaxy clusters Abell 2667 and 2390, using spectra from VIMOS.

\begin{figure}
    \centering
    \includegraphics[width=0.48\textwidth]{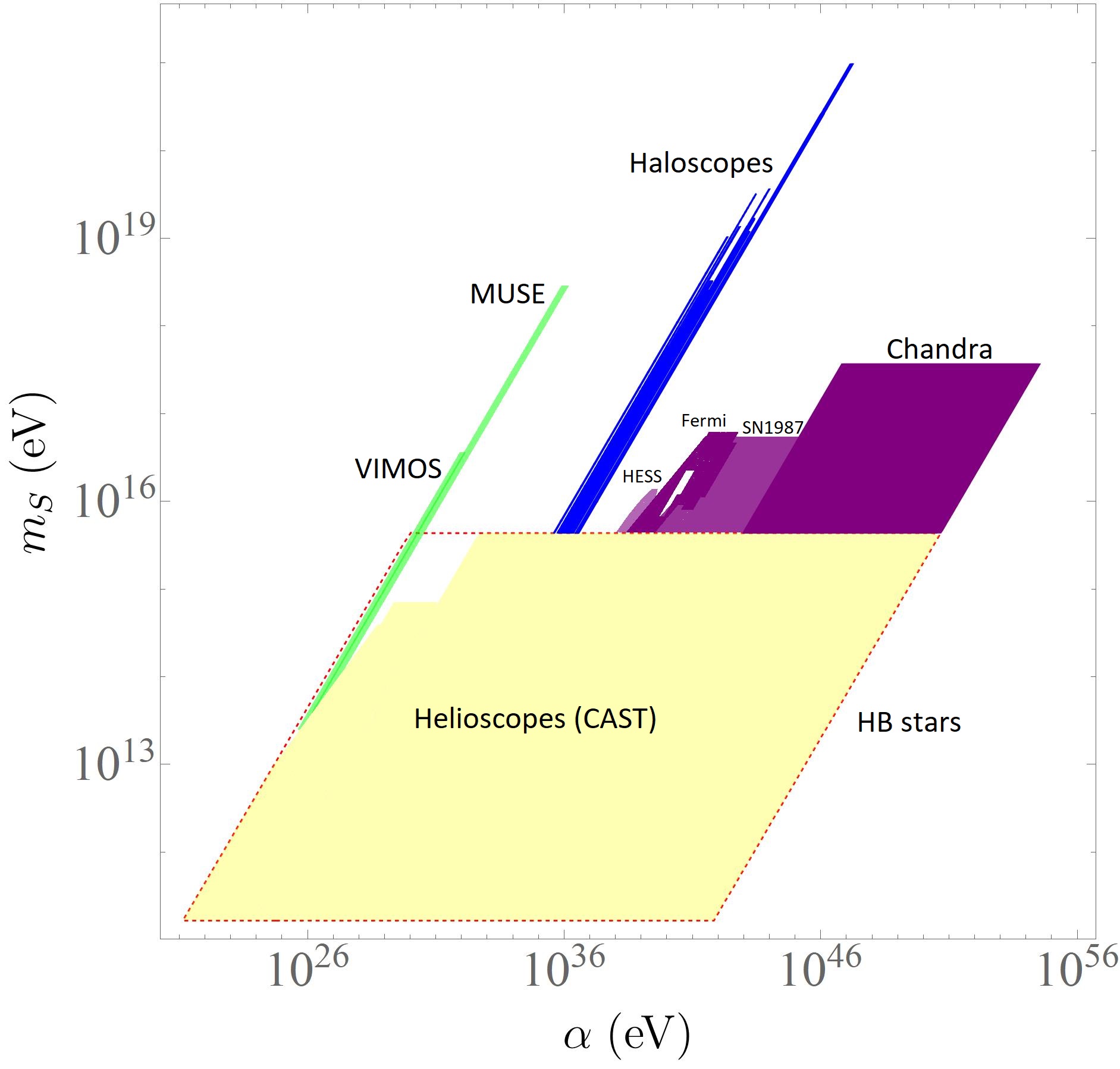}
    \caption{Exclusion plot in the $m_S-\alpha$ parameter space corresponding to constraints on ALPs based on the coupling to the electromagnetic sector. We have also taken into account that the $m_S<0.166{\rm{~TeV}}$ from the four-fermion contact interaction. 
}
    \label{fig:2}
\end{figure}

A representation of the constraints above and the pertinent references to the experiments can be found in \cite{Semertzidis:2021rxs}. By using the relation between $g_{\phi\gamma\gamma}$ and $m_S$, it can be easily seen that the limits imposed on $m_S$ in these studies turn to be stronger than the ones from the four-fermion contact interaction. We represent the experimental constraints derived from the electromagnetic coupling in the $m_S-\alpha$ parameter space in Figure \ref{fig:2}.

4) The coupling of ALPs to fermions, \emph{i.e.}, $g_{\phi ff}\Delta_{\rm N}^{\mu}\partial_{\mu}\phi$, produce spin flips in a magnetic sample placed inside a static magnetic field. Such spin flips would then emit radio-frequency photons that can be detected by a suitable quantum counter in an ultra-cryogenic environment \cite{Barbieri:2016vwg}. Based on this effect,the QUAX experiments has put the strongest constraints on the strength of this coupling \cite{QUAX:2020adt}. In particular, a constraint of $g_{\phi ff}<1.66\cdot10^{-5}\:{\rm{TeV}}^{-1}$ is found for masses in the interval $42.4-43.1 \:\mu \mathrm{eV}$. This translates in a lower limit of $6.02\cdot 10^{4}\:{\rm{TeV}}$ for $m_S$ for the mentioned ALP mass range.

Finally, let us focus on ALPs constraints for ultralight and heavy masses. On the one hand, recent analysis of the Lyman-alpha forest searching for suppressed cosmic structure growth, have shown a lower limit on the mass of ultra-light axions of $2\cdot 10^{-20}\:{\rm{eV}}$ \cite{Rogers:2020ltq}. On the other hand, the upper limits on the mass of the particle are found based on decays to SM particles. Such a decay affects the abundance of light elements in the Universe, and hence it may affect the cosmological evolution. The ALP coupling with the electromagnetic sector allows ALPs to decay into two photons, with a lifetime \cite{Marsh:2015xka}
\begin{equation}
    \tau_{\phi\gamma\gamma}=\frac{4\pi}{m_{\phi}^{3}g_{\phi\gamma\gamma}^{2}}\approx71.87\:{\rm s}\left(\frac{{\rm TeV}}{m_{S}}\right)^{4}\left(\frac{\alpha}{10^{7}\:{\rm TeV}}\right)^{3}.
\end{equation}

\begin{figure}
    \centering
    \includegraphics[width=0.48\textwidth]{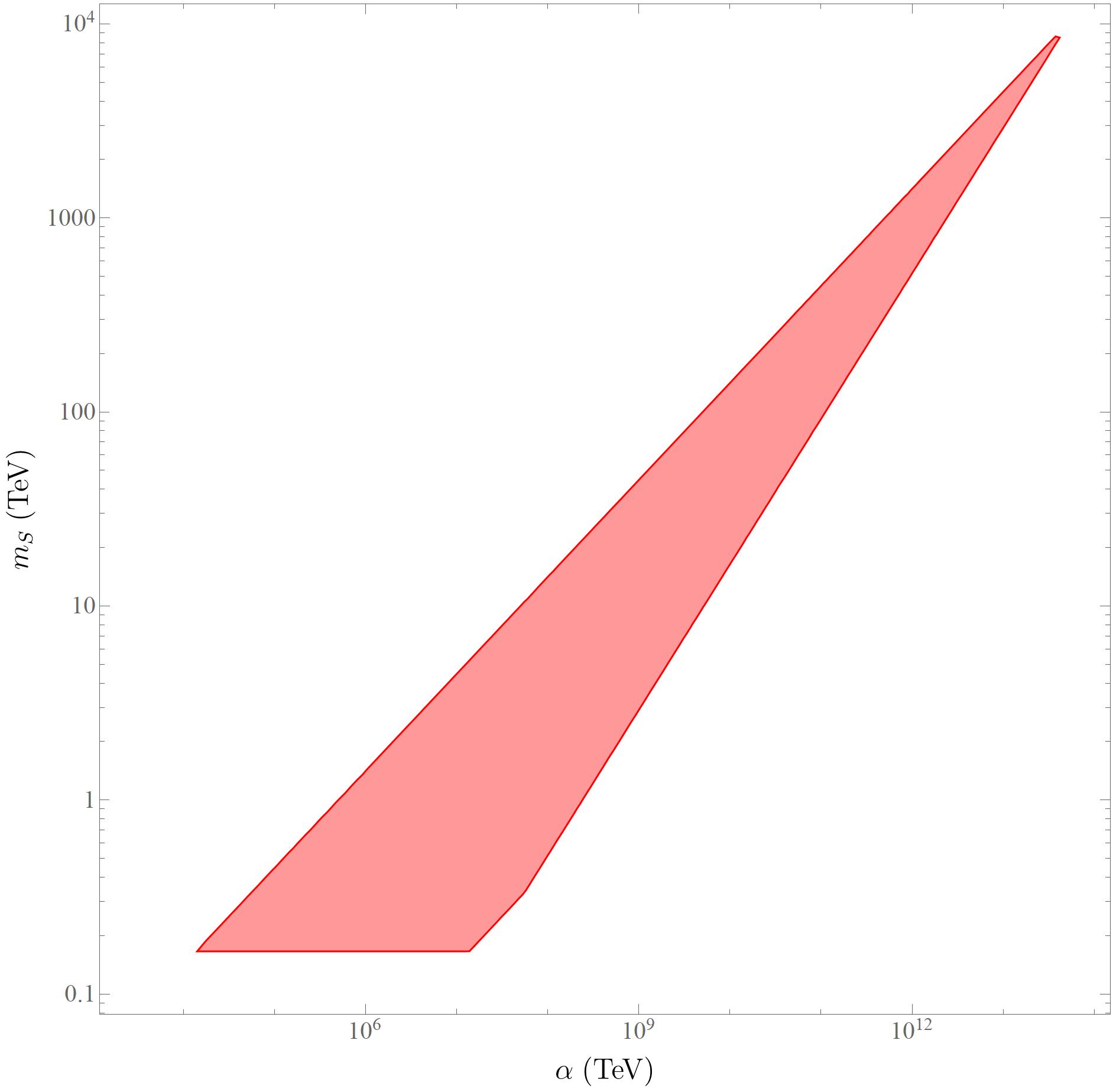}
    \caption{Exclusion plot in the $m_S-\alpha$ parameter space corresponding to the constraints on ALPs on the keV-GeV range based on the lifetime of the ALP. We also take into account the $m_S<0.166\:{\rm{TeV}}$ from the four-fermion contact interaction. 
}
    \label{fig:1}
\end{figure}

If $\tau_{\phi\gamma\gamma}\gtrsim\tau_{{\rm univ}}$, where $\tau_{{\rm univ}}$ is the age of the Universe, then the ALP would be stable for the lifetime of the Universe. In the $m_S-\alpha$ parameter space, such a condition translates to $\alpha\left[{\rm TeV}\right]\gtrsim1.82\cdot10^{12}\left(\frac{m_{S}}{{\rm TeV}}\right)^{\frac{4}{3}}$. Otherwise, the decay of the ALP would affect the cosmological evolution, and hence constraints can be imposed. In particular, it is found that the ALPs with the following range of masses and lifetimes are excluded \cite{Marsh:2015xka}:
\begin{equation}
    1\:{\rm keV}\lesssim m_{\phi}\lesssim1\:{\rm GeV},\quad10^{-4}\mathrm{~s}\lesssim\tau_{\phi\gamma\gamma}\lesssim10^{6}\mathrm{~s}.
\end{equation}
Such constraints have been translated to the $m_S-\alpha$ parameter space in Figure \ref{fig:1}.\\

\textit{Describing {\rm CDM}.}--- As we have established, the Poincar\'e Gauge gravity pseudo-scalar d.o.f. is an ALP. Provided that the experimental constraints are fulfilled, this predicted particle could serve as a DM candidate. Nevertheless, we have to make sure that the early Universe production of such a particle can account for the DM density measured by observations \cite{Planck:2018vyg}. 
In order to perform the density calculation we assume that the production of the ALP is done via the misalignment mechanism. Such a mechanism relies on the fact that fields in the early Universe have a random initial state. After the mass of the particle is comparable to the Hubble parameter, the fields respond by attempting to minimise their potential, hence oscillating around the minimum. These oscillations can behave as CDM since their energy density is diluted by the expansion of the Universe as $\rho\sim a^{-3}$ \cite{Arias:2012az}.
After studying the pseudo-scalar evolution in a Friedman-Lemaître-Robertson-Walker (FLRW) background, one can show that the production mechanism starts when $m_{\phi}\simeq3H\left(T\right)$. Hence, the relation between the Hubble parameter $H$ and the temperature $T$ in the radiation dominated era can be used to find the temperature at which the mechanism starts. %Such a relation is given by \cite{Kolb:1990vq}\begin{equation}H\left(T\right)=1.66\frac{\sqrt{g_{*}\left(T\right)}T^{2}}{m_{Pl}},\end{equation}
Thus, knowing the temperature at which the mechanism begins to take place $T_{\rm{mis}}$, we can calculate the expected density today from this production as \cite{Arias:2012az}
\begin{equation}
    \Omega_{\phi}h^{2}=2.542\sqrt{\frac{m_{\phi}}{{\rm eV}}}\left(\frac{\phi_{{\rm mis}}}{10^{18}\mathrm{~ eV}}\right)\frac{g_{*}\left(T_{{\rm mis}}\right)^{3/4}}{g_{*S}\left(T_{{\rm mis}}\right)},
\end{equation}
where 
%$m_{Pl}$ is the Planck mass and 
$g_{*}$ and $g_{*S}$ are the effective numbers of energy and entropy d.o.f.s respectively \cite{Husdal:2016haj}. Moreover, we have taken into account that in this case the ALP mass is constant, and we have used the value $\rho_{{\rm crit}}=1.053672\,h^{2}\mathrm{~GeV/cm^{3}}$ for the critical density of the Universe \cite{ParticleDataGroup:2020ssz}.

\begin{figure}
\begin{subfigure}{.5\linewidth}
\centering
\includegraphics[width=1\linewidth]{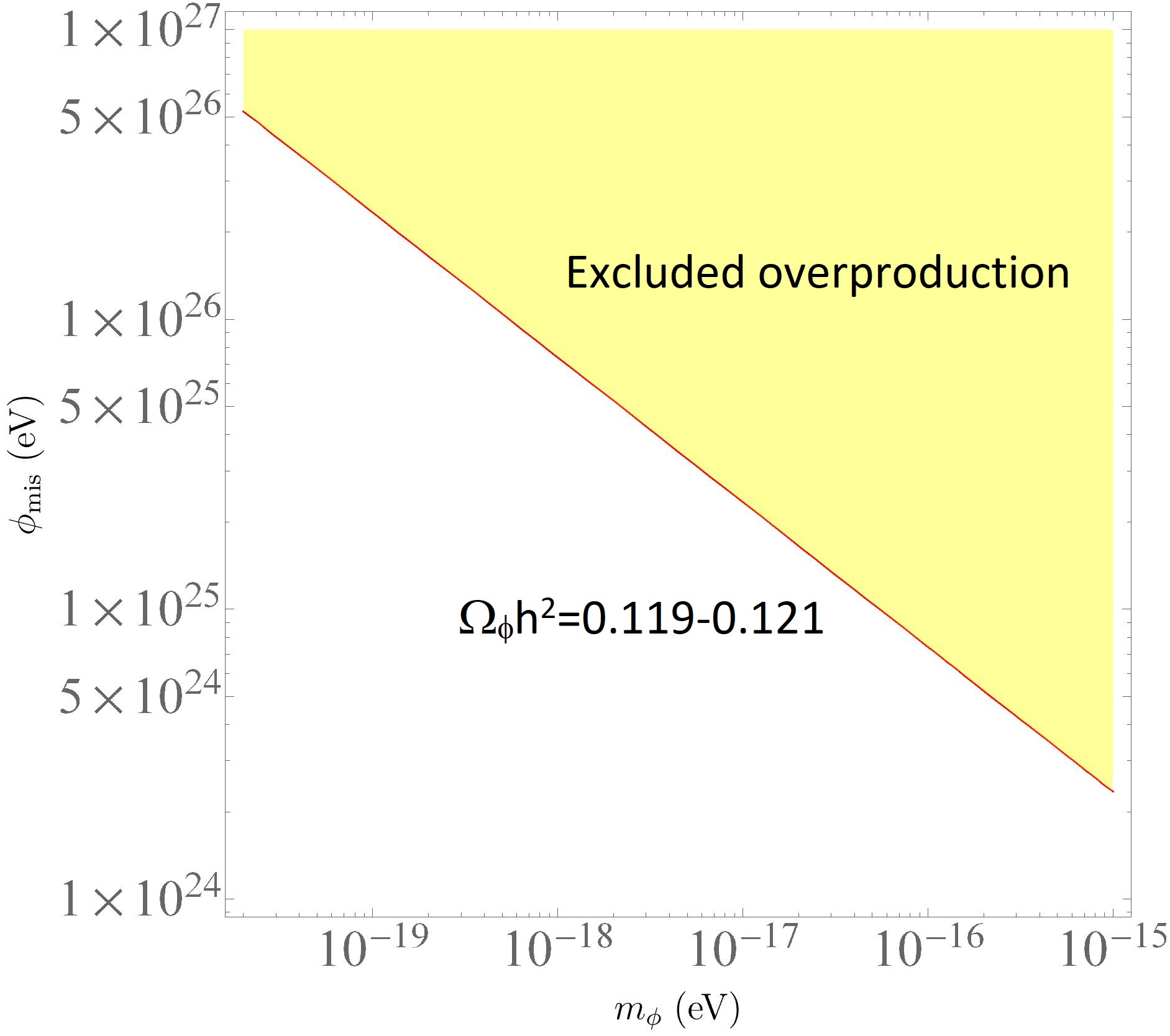}
\caption{$2\cdot10^{-20}\mathrm{~eV}\leq m_{\phi}\leq10^{-15}\mathrm{~eV}$}
\label{fig31}
\end{subfigure}%
\begin{subfigure}{.5\linewidth}
\centering
\includegraphics[width=1\linewidth]{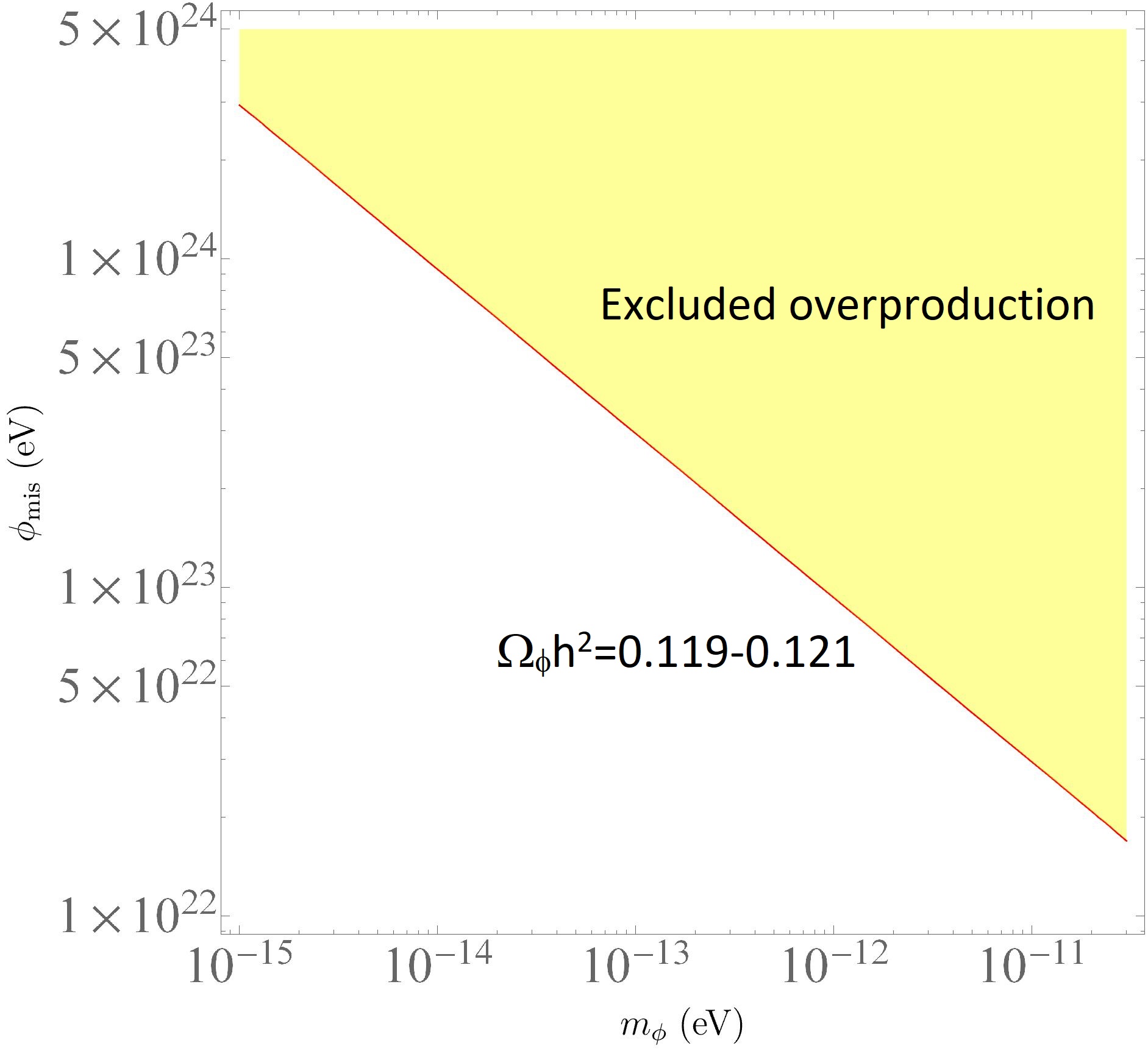}
\caption{$10^{-15}\mathrm{~eV}<m_{\phi}\leq3\cdot10^{-11}\mathrm{~eV}$}
\label{fig32}
\end{subfigure}\\[1ex]
\begin{subfigure}{\linewidth}
\centering
\includegraphics[width=0.5\linewidth]{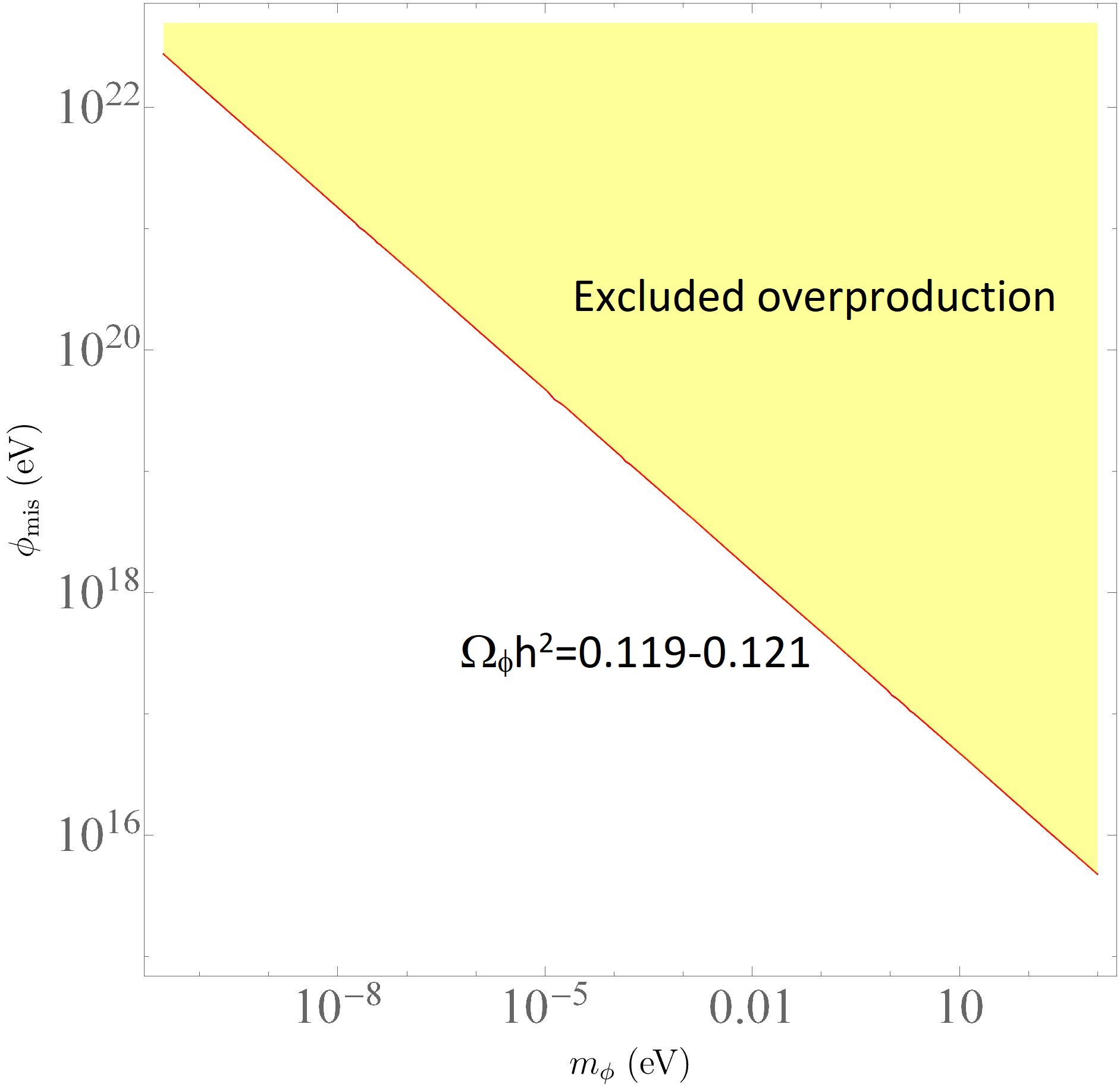}
\caption{$m_{\phi}>3\cdot10^{-11}\mathrm{eV}$}
\label{fig33}
\end{subfigure}
\caption{Exclusion plots in the $m_{\phi}-\phi_{\rm{mis}}$ parameter space corresponding to the constraints on ALPs based on the predicted dark matter density from the misalignment mechanism. The red line represents the allowed values able to account for the whole dark matter and the yellow region represents the excluded parameter space by dark matter overproduction. We asssume $g_{*}$ and $g_{*S}$ to be: $g_{*}=3.363\:,\:g_{*S}=3.909$ for \eqref{fig31}; $g_{*}=g_{*S}=15.25$ for \eqref{fig32}; $g_{*}=g_{*S}=106.75$ for \eqref{fig33} \cite{Husdal:2016haj}.}
\label{fig:3}
\end{figure}

The most precise value for the CDM density today, $\Omega_{\rm{CDM}}h^{2}=0.120\pm 0.001$, is given by the Planck Collaboration \cite{Planck:2018vyg}. Hence, if we want the ALP to describe the entire dark matter content, we shall require $\Omega_{\phi}=\Omega_{\rm{CDM}}$. This implies that the initial value of the field after inflation imposes a very limited allowed mass range, and vice versa. Also, the values of $m_{\phi}$ and $\phi_{\rm{mis}}$ providing higher densities of dark matter would be of course excluded. These constraints are represented in Figure \ref{fig:3}. 
Furthermore, the constraints on $\phi_{\rm{mis}}$ would affect the possible values of $m_T$ if we require the weak-field limit approximation to be valid. Such an approximation requires that $m_T\gtrsim 10\,\phi_{\rm{mis}}$. Hence, from Figure \ref{fig:3} we can infer the possible values of $m_T$ such that the pseudo-scalar accounts for the whole cold dark matter while the weak-field limit is still applicable.\\

In conclusion, we show that the pseudo-scalar mode of Poincar\'e Gauge Gravity behaves like an axion-like particle. We find experimental constraints for the free parameters from the interactions present in the theory. Finally, we provide the  conditions for which the pseudo-scalar mode present in this theory can account for the entirety of cold dark matter. \\

\begin{acknowledgments}
We thank J. Beltr\'an Jim\'enez, J.A.R. Cembranos, and J. Gigante Valcarcel for helpful conversations. FJMT and AdlCD acknowledge  support from NRF grants no.120390, reference:BSFP190416431035; no.120396, reference:CSRP190405427545; no.101775, reference: SFH150727131568. AdlCD acknowledges  support from grants PID2019-108655GB-I00 and
COOPB204064, I-COOP+2019, MICINN Spain.
DFM and FJMT acknowledge support from the Research Council of Norway.
\end{acknowledgments}
\bibliographystyle{apsrev}
\bibliography{sorsamp}% Produces the bibliography via BibTeX.

\providecommand{\noopsort}[1]{}\providecommand{\singleletter}[1]{#1}%
\begin{thebibliography}{39}
\expandafter\ifx\csname natexlab\endcsname\relax\def\natexlab#1{#1}\fi
\expandafter\ifx\csname bibnamefont\endcsname\relax
  \def\bibnamefont#1{#1}\fi
\expandafter\ifx\csname bibfnamefont\endcsname\relax
  \def\bibfnamefont#1{#1}\fi
\expandafter\ifx\csname citenamefont\endcsname\relax
  \def\citenamefont#1{#1}\fi
\expandafter\ifx\csname url\endcsname\relax
  \def\url#1{\texttt{#1}}\fi
\expandafter\ifx\csname urlprefix\endcsname\relax\def\urlprefix{URL }\fi
\providecommand{\bibinfo}[2]{#2}
\providecommand{\eprint}[2][]{\url{#2}}

\bibitem[{\citenamefont{Riess et~al.}(1998)}]{Riess:1998cb}
\bibinfo{author}{\bibfnamefont{A.~G.} \bibnamefont{Riess}} \bibnamefont{et~al.}
  (\bibinfo{collaboration}{Supernova Search Team}), \bibinfo{journal}{Astron.
  J.} \textbf{\bibinfo{volume}{116}}, \bibinfo{pages}{1009}
  (\bibinfo{year}{1998}).

\bibitem[{\citenamefont{Sahni}(2004)}]{Sahni:2004ai}
\bibinfo{author}{\bibfnamefont{V.}~\bibnamefont{Sahni}},
  \bibinfo{journal}{Lect. Notes Phys.} \textbf{\bibinfo{volume}{653}},
  \bibinfo{pages}{141} (\bibinfo{year}{2004}), \eprint{astro-ph/0403324}.

\bibitem[{\citenamefont{Peebles and Ratra}(2003)}]{Peebles:2002gy}
\bibinfo{author}{\bibfnamefont{P.~J.~E.} \bibnamefont{Peebles}}
  \bibnamefont{and} \bibinfo{author}{\bibfnamefont{B.}~\bibnamefont{Ratra}},
  \bibinfo{journal}{Rev. Mod. Phys.} \textbf{\bibinfo{volume}{75}},
  \bibinfo{pages}{559} (\bibinfo{year}{2003}), \eprint{astro-ph/0207347}.

\bibitem[{\citenamefont{Weinberg}(1989)}]{Weinberg:1988cp}
\bibinfo{author}{\bibfnamefont{S.}~\bibnamefont{Weinberg}},
  \bibinfo{journal}{Rev. Mod. Phys.} \textbf{\bibinfo{volume}{61}},
  \bibinfo{pages}{1} (\bibinfo{year}{1989}).

\bibitem[{\citenamefont{Bertone et~al.}(2005)\citenamefont{Bertone, Hooper, and
  Silk}}]{Bertone:2004pz}
\bibinfo{author}{\bibfnamefont{G.}~\bibnamefont{Bertone}},
  \bibinfo{author}{\bibfnamefont{D.}~\bibnamefont{Hooper}}, \bibnamefont{and}
  \bibinfo{author}{\bibfnamefont{J.}~\bibnamefont{Silk}},
  \bibinfo{journal}{Phys. Rept.} \textbf{\bibinfo{volume}{405}},
  \bibinfo{pages}{279} (\bibinfo{year}{2005}), \eprint{hep-ph/0404175}.

\bibitem[{\citenamefont{Cembranos}(2009)}]{Cembranos:2008gj}
\bibinfo{author}{\bibfnamefont{J.~A.~R.} \bibnamefont{Cembranos}},
  \bibinfo{journal}{Phys. Rev. Lett.} \textbf{\bibinfo{volume}{102}},
  \bibinfo{pages}{141301} (\bibinfo{year}{2009}), \eprint{0809.1653}.

\bibitem[{\citenamefont{Clifton et~al.}(2012)\citenamefont{Clifton, Ferreira,
  Padilla, and Skordis}}]{Clifton:2011jh}
\bibinfo{author}{\bibfnamefont{T.}~\bibnamefont{Clifton}},
  \bibinfo{author}{\bibfnamefont{P.~G.} \bibnamefont{Ferreira}},
  \bibinfo{author}{\bibfnamefont{A.}~\bibnamefont{Padilla}}, \bibnamefont{and}
  \bibinfo{author}{\bibfnamefont{C.}~\bibnamefont{Skordis}},
  \bibinfo{journal}{Phys. Rept.} \textbf{\bibinfo{volume}{513}},
  \bibinfo{pages}{1} (\bibinfo{year}{2012}), \eprint{1106.2476}.

\bibitem[{\citenamefont{Amendola et~al.}(2018)}]{Amendola:2016saw}
\bibinfo{author}{\bibfnamefont{L.}~\bibnamefont{Amendola}}
  \bibnamefont{et~al.}, \bibinfo{journal}{Living Rev. Rel.}
  \textbf{\bibinfo{volume}{21}}, \bibinfo{pages}{2} (\bibinfo{year}{2018}),
  \eprint{1606.00180}.

\bibitem[{\citenamefont{Blagojevi\'c and Hehl}(2013)}]{Blagojevic:2013xpa}
\bibinfo{editor}{\bibfnamefont{M.}~\bibnamefont{Blagojevi\'c}}
  \bibnamefont{and} \bibinfo{editor}{\bibfnamefont{F.~W.} \bibnamefont{Hehl}},
  eds., \emph{\bibinfo{title}{{Gauge Theories of Gravitation}: {A Reader with
  Commentaries}}} (\bibinfo{publisher}{World Scientific},
  \bibinfo{address}{Singapore}, \bibinfo{year}{2013}), ISBN
  \bibinfo{isbn}{978-1-84816-726-1}.

\bibitem[{\citenamefont{Ponomarev et~al.}(2017)\citenamefont{Ponomarev,
  Obukhov, and Barvinsky}}]{ponomarev2017gauge}
\bibinfo{author}{\bibfnamefont{V.~N.} \bibnamefont{Ponomarev}},
  \bibinfo{author}{\bibfnamefont{Y.~V.} \bibnamefont{Obukhov}},
  \bibnamefont{and}
  \bibinfo{author}{\bibfnamefont{A.}~\bibnamefont{Barvinsky}},
  \emph{\bibinfo{title}{Gauge approach and quantization methods in gravity
  theory}} (\bibinfo{publisher}{Nauka}, \bibinfo{year}{2017}).

\bibitem[{\citenamefont{Hayashi and Shirafuji}(1980)}]{Hayashi:1980qp}
\bibinfo{author}{\bibfnamefont{K.}~\bibnamefont{Hayashi}} \bibnamefont{and}
  \bibinfo{author}{\bibfnamefont{T.}~\bibnamefont{Shirafuji}},
  \bibinfo{journal}{Prog.\ Theor.\ Phys.} \textbf{\bibinfo{volume}{64}},
  \bibinfo{pages}{2222} (\bibinfo{year}{1980}).

\bibitem[{\citenamefont{Yo and Nester}(1999)}]{Yo:1999ex}
\bibinfo{author}{\bibfnamefont{H.-j.} \bibnamefont{Yo}} \bibnamefont{and}
  \bibinfo{author}{\bibfnamefont{J.~M.} \bibnamefont{Nester}},
  \bibinfo{journal}{Int. J. Mod. Phys. D} \textbf{\bibinfo{volume}{8}},
  \bibinfo{pages}{459} (\bibinfo{year}{1999}), \eprint{gr-qc/9902032}.

\bibitem[{\citenamefont{Yo and Nester}(2002)}]{Yo:2001sy}
\bibinfo{author}{\bibfnamefont{H.-J.} \bibnamefont{Yo}} \bibnamefont{and}
  \bibinfo{author}{\bibfnamefont{J.~M.} \bibnamefont{Nester}},
  \bibinfo{journal}{Int. J. Mod. Phys. D} \textbf{\bibinfo{volume}{11}},
  \bibinfo{pages}{747} (\bibinfo{year}{2002}), \eprint{gr-qc/0112030}.

\bibitem[{\citenamefont{Jim\'enez and
  Maldonado~Torralba}(2020)}]{Jimenez:2019qjc}
\bibinfo{author}{\bibfnamefont{J.~B.} \bibnamefont{Jim\'enez}}
  \bibnamefont{and} \bibinfo{author}{\bibfnamefont{F.~J.}
  \bibnamefont{Maldonado~Torralba}}, \bibinfo{journal}{Eur. Phys. J. C}
  \textbf{\bibinfo{volume}{80}}, \bibinfo{pages}{611} (\bibinfo{year}{2020}),
  \eprint{1910.07506}.

\bibitem[{\citenamefont{Hojman et~al.}(1980)\citenamefont{Hojman, Mukku, and
  Sayed}}]{Hojman:1980kv}
\bibinfo{author}{\bibfnamefont{R.}~\bibnamefont{Hojman}},
  \bibinfo{author}{\bibfnamefont{C.}~\bibnamefont{Mukku}}, \bibnamefont{and}
  \bibinfo{author}{\bibfnamefont{W.}~\bibnamefont{Sayed}},
  \bibinfo{journal}{Phys. Rev. D} \textbf{\bibinfo{volume}{22}},
  \bibinfo{pages}{1915} (\bibinfo{year}{1980}).

\bibitem[{\citenamefont{Holst}(1996)}]{Holst:1995pc}
\bibinfo{author}{\bibfnamefont{S.}~\bibnamefont{Holst}},
  \bibinfo{journal}{Phys. Rev.} \textbf{\bibinfo{volume}{D53}},
  \bibinfo{pages}{5966} (\bibinfo{year}{1996}), \eprint{gr-qc/9511026}.

\bibitem[{\citenamefont{Castellani et~al.}(1991)\citenamefont{Castellani,
  D'auria, and Fre}}]{castellani1991supergravity}
\bibinfo{author}{\bibfnamefont{L.}~\bibnamefont{Castellani}},
  \bibinfo{author}{\bibfnamefont{R.}~\bibnamefont{D'auria}}, \bibnamefont{and}
  \bibinfo{author}{\bibfnamefont{P.}~\bibnamefont{Fre}},
  \emph{\bibinfo{title}{Supergravity and superstrings: a geometric perspective
  (in 3 volumes)}}, vol.~\bibinfo{volume}{1} (\bibinfo{publisher}{World
  Scientific Publishing Company}, \bibinfo{year}{1991}).

\bibitem[{\citenamefont{Taveras and Yunes}(2008)}]{Taveras:2008yf}
\bibinfo{author}{\bibfnamefont{V.}~\bibnamefont{Taveras}} \bibnamefont{and}
  \bibinfo{author}{\bibfnamefont{N.}~\bibnamefont{Yunes}},
  \bibinfo{journal}{Phys. Rev.} \textbf{\bibinfo{volume}{D78}},
  \bibinfo{pages}{064070} (\bibinfo{year}{2008}), \eprint{0807.2652}.

\bibitem[{\citenamefont{Calcagni and Mercuri}(2009)}]{Calcagni:2009xz}
\bibinfo{author}{\bibfnamefont{G.}~\bibnamefont{Calcagni}} \bibnamefont{and}
  \bibinfo{author}{\bibfnamefont{S.}~\bibnamefont{Mercuri}},
  \bibinfo{journal}{Phys. Rev.} \textbf{\bibinfo{volume}{D79}},
  \bibinfo{pages}{084004} (\bibinfo{year}{2009}), \eprint{0902.0957}.

\bibitem[{\citenamefont{Torres-Gomez and Krasnov}(2009)}]{TorresGomez:2008fj}
\bibinfo{author}{\bibfnamefont{A.}~\bibnamefont{Torres-Gomez}}
  \bibnamefont{and} \bibinfo{author}{\bibfnamefont{K.}~\bibnamefont{Krasnov}},
  \bibinfo{journal}{Phys. Rev. D} \textbf{\bibinfo{volume}{79}},
  \bibinfo{pages}{104014} (\bibinfo{year}{2009}), \eprint{0811.1998}.

\bibitem[{\citenamefont{Mercuri}(2009)}]{Mercuri:2009zi}
\bibinfo{author}{\bibfnamefont{S.}~\bibnamefont{Mercuri}},
  \bibinfo{journal}{Phys. Rev. Lett.} \textbf{\bibinfo{volume}{103}},
  \bibinfo{pages}{081302} (\bibinfo{year}{2009}), \eprint{0902.2764}.

\bibitem[{\citenamefont{Shapiro}(2002)}]{Shapiro:2001rz}
\bibinfo{author}{\bibfnamefont{I.~L.} \bibnamefont{Shapiro}},
  \bibinfo{journal}{Phys. Rept.} \textbf{\bibinfo{volume}{357}},
  \bibinfo{pages}{113} (\bibinfo{year}{2002}), \eprint{hep-th/0103093}.

\bibitem[{\citenamefont{Zyla et~al.}(2020)}]{ParticleDataGroup:2020ssz}
\bibinfo{author}{\bibfnamefont{P.~A.} \bibnamefont{Zyla}} \bibnamefont{et~al.}
  (\bibinfo{collaboration}{Particle Data Group}), \bibinfo{journal}{PTEP}
  \textbf{\bibinfo{volume}{2020}}, \bibinfo{pages}{083C01}
  (\bibinfo{year}{2020}).

\bibitem[{\citenamefont{Marsh}(2016)}]{Marsh:2015xka}
\bibinfo{author}{\bibfnamefont{D.~J.~E.} \bibnamefont{Marsh}},
  \bibinfo{journal}{Phys. Rept.} \textbf{\bibinfo{volume}{643}},
  \bibinfo{pages}{1} (\bibinfo{year}{2016}), \eprint{1510.07633}.

\bibitem[{\citenamefont{Bertlmann}(1996)}]{Bertlmann:1996xk}
\bibinfo{author}{\bibfnamefont{R.~A.} \bibnamefont{Bertlmann}},
  \emph{\bibinfo{title}{{Anomalies in quantum field theory}}}
  (\bibinfo{year}{1996}).

\bibitem[{\citenamefont{Zarnecki}(1999)}]{Zarnecki:1999je}
\bibinfo{author}{\bibfnamefont{A.~F.} \bibnamefont{Zarnecki}},
  \bibinfo{journal}{Eur. Phys. J. C} \textbf{\bibinfo{volume}{11}},
  \bibinfo{pages}{539} (\bibinfo{year}{1999}), \eprint{hep-ph/9904334}.

\bibitem[{\citenamefont{Alexander and Yunes}(2009)}]{Alexander:2009tp}
\bibinfo{author}{\bibfnamefont{S.}~\bibnamefont{Alexander}} \bibnamefont{and}
  \bibinfo{author}{\bibfnamefont{N.}~\bibnamefont{Yunes}},
  \bibinfo{journal}{Phys. Rept.} \textbf{\bibinfo{volume}{480}},
  \bibinfo{pages}{1} (\bibinfo{year}{2009}), \eprint{0907.2562}.

\bibitem[{\citenamefont{Jung et~al.}(2020)\citenamefont{Jung, Kim, Soda, and
  Urakawa}}]{Jung:2020aem}
\bibinfo{author}{\bibfnamefont{S.}~\bibnamefont{Jung}},
  \bibinfo{author}{\bibfnamefont{T.}~\bibnamefont{Kim}},
  \bibinfo{author}{\bibfnamefont{J.}~\bibnamefont{Soda}}, \bibnamefont{and}
  \bibinfo{author}{\bibfnamefont{Y.}~\bibnamefont{Urakawa}},
  \bibinfo{journal}{Phys. Rev. D} \textbf{\bibinfo{volume}{102}},
  \bibinfo{pages}{055013} (\bibinfo{year}{2020}), \eprint{2003.02853}.

\bibitem[{\citenamefont{Yunes and Pretorius}(2009)}]{Yunes:2009hc}
\bibinfo{author}{\bibfnamefont{N.}~\bibnamefont{Yunes}} \bibnamefont{and}
  \bibinfo{author}{\bibfnamefont{F.}~\bibnamefont{Pretorius}},
  \bibinfo{journal}{Phys. Rev. D} \textbf{\bibinfo{volume}{79}},
  \bibinfo{pages}{084043} (\bibinfo{year}{2009}), \eprint{0902.4669}.

\bibitem[{\citenamefont{Ali-Haimoud and Chen}(2011)}]{Ali-Haimoud:2011zme}
\bibinfo{author}{\bibfnamefont{Y.}~\bibnamefont{Ali-Haimoud}} \bibnamefont{and}
  \bibinfo{author}{\bibfnamefont{Y.}~\bibnamefont{Chen}},
  \bibinfo{journal}{Phys. Rev. D} \textbf{\bibinfo{volume}{84}},
  \bibinfo{pages}{124033} (\bibinfo{year}{2011}), \eprint{1110.5329}.

\bibitem[{\citenamefont{Chadha-Day et~al.}(2021)\citenamefont{Chadha-Day,
  Ellis, and Marsh}}]{Chadha-Day:2021szb}
\bibinfo{author}{\bibfnamefont{F.}~\bibnamefont{Chadha-Day}},
  \bibinfo{author}{\bibfnamefont{J.}~\bibnamefont{Ellis}}, \bibnamefont{and}
  \bibinfo{author}{\bibfnamefont{D.~J.~E.} \bibnamefont{Marsh}},
  \emph{\bibinfo{title}{{Axion Dark Matter: What is it and Why Now?}}}
  (\bibinfo{year}{2021}), \eprint{2105.01406}.

\bibitem[{\citenamefont{Semertzidis and Youn}(2021)}]{Semertzidis:2021rxs}
\bibinfo{author}{\bibfnamefont{Y.~K.} \bibnamefont{Semertzidis}}
  \bibnamefont{and} \bibinfo{author}{\bibfnamefont{S.}~\bibnamefont{Youn}},
  \emph{\bibinfo{title}{{Axion Dark Matter: How to detect it?}}}
  (\bibinfo{year}{2021}), \eprint{2104.14831}.

\bibitem[{\citenamefont{Ayala et~al.}(2014)\citenamefont{Ayala, Dom\'\i{}nguez,
  Giannotti, Mirizzi, and Straniero}}]{Ayala:2014pea}
\bibinfo{author}{\bibfnamefont{A.}~\bibnamefont{Ayala}},
  \bibinfo{author}{\bibfnamefont{I.}~\bibnamefont{Dom\'\i{}nguez}},
  \bibinfo{author}{\bibfnamefont{M.}~\bibnamefont{Giannotti}},
  \bibinfo{author}{\bibfnamefont{A.}~\bibnamefont{Mirizzi}}, \bibnamefont{and}
  \bibinfo{author}{\bibfnamefont{O.}~\bibnamefont{Straniero}},
  \bibinfo{journal}{Phys. Rev. Lett.} \textbf{\bibinfo{volume}{113}},
  \bibinfo{pages}{191302} (\bibinfo{year}{2014}), \eprint{1406.6053}.

\bibitem[{\citenamefont{Barbieri et~al.}(2017)\citenamefont{Barbieri, Braggio,
  Carugno, Gallo, Lombardi, Ortolan, Pengo, Ruoso, and
  Speake}}]{Barbieri:2016vwg}
\bibinfo{author}{\bibfnamefont{R.}~\bibnamefont{Barbieri}},
  \bibinfo{author}{\bibfnamefont{C.}~\bibnamefont{Braggio}},
  \bibinfo{author}{\bibfnamefont{G.}~\bibnamefont{Carugno}},
  \bibinfo{author}{\bibfnamefont{C.~S.} \bibnamefont{Gallo}},
  \bibinfo{author}{\bibfnamefont{A.}~\bibnamefont{Lombardi}},
  \bibinfo{author}{\bibfnamefont{A.}~\bibnamefont{Ortolan}},
  \bibinfo{author}{\bibfnamefont{R.}~\bibnamefont{Pengo}},
  \bibinfo{author}{\bibfnamefont{G.}~\bibnamefont{Ruoso}}, \bibnamefont{and}
  \bibinfo{author}{\bibfnamefont{C.~C.} \bibnamefont{Speake}},
  \bibinfo{journal}{Phys. Dark Univ.} \textbf{\bibinfo{volume}{15}},
  \bibinfo{pages}{135} (\bibinfo{year}{2017}), \eprint{1606.02201}.

\bibitem[{\citenamefont{Crescini et~al.}(2020)}]{QUAX:2020adt}
\bibinfo{author}{\bibfnamefont{N.}~\bibnamefont{Crescini}} \bibnamefont{et~al.}
  (\bibinfo{collaboration}{QUAX}), \bibinfo{journal}{Phys. Rev. Lett.}
  \textbf{\bibinfo{volume}{124}}, \bibinfo{pages}{171801}
  (\bibinfo{year}{2020}), \eprint{2001.08940}.

\bibitem[{\citenamefont{Rogers and Peiris}(2021)}]{Rogers:2020ltq}
\bibinfo{author}{\bibfnamefont{K.~K.} \bibnamefont{Rogers}} \bibnamefont{and}
  \bibinfo{author}{\bibfnamefont{H.~V.} \bibnamefont{Peiris}},
  \bibinfo{journal}{Phys. Rev. Lett.} \textbf{\bibinfo{volume}{126}},
  \bibinfo{pages}{071302} (\bibinfo{year}{2021}), \eprint{2007.12705}.

\bibitem[{\citenamefont{Aghanim et~al.}(2020)}]{Planck:2018vyg}
\bibinfo{author}{\bibfnamefont{N.}~\bibnamefont{Aghanim}} \bibnamefont{et~al.}
  (\bibinfo{collaboration}{Planck}), \bibinfo{journal}{Astron. Astrophys.}
  \textbf{\bibinfo{volume}{641}}, \bibinfo{pages}{A6} (\bibinfo{year}{2020}),
  \bibinfo{note}{[Erratum: Astron.Astrophys. 652, C4 (2021)]},
  \eprint{1807.06209}.

\bibitem[{\citenamefont{Arias et~al.}(2012)\citenamefont{Arias, Cadamuro,
  Goodsell, Jaeckel, Redondo, and Ringwald}}]{Arias:2012az}
\bibinfo{author}{\bibfnamefont{P.}~\bibnamefont{Arias}},
  \bibinfo{author}{\bibfnamefont{D.}~\bibnamefont{Cadamuro}},
  \bibinfo{author}{\bibfnamefont{M.}~\bibnamefont{Goodsell}},
  \bibinfo{author}{\bibfnamefont{J.}~\bibnamefont{Jaeckel}},
  \bibinfo{author}{\bibfnamefont{J.}~\bibnamefont{Redondo}}, \bibnamefont{and}
  \bibinfo{author}{\bibfnamefont{A.}~\bibnamefont{Ringwald}},
  \bibinfo{journal}{JCAP} \textbf{\bibinfo{volume}{06}}, \bibinfo{pages}{013}
  (\bibinfo{year}{2012}), \eprint{1201.5902}.

\bibitem[{\citenamefont{Husdal}(2016)}]{Husdal:2016haj}
\bibinfo{author}{\bibfnamefont{L.}~\bibnamefont{Husdal}},
  \bibinfo{journal}{Galaxies} \textbf{\bibinfo{volume}{4}}, \bibinfo{pages}{78}
  (\bibinfo{year}{2016}), \eprint{1609.04979}.

\end{thebibliography}
\end{document}